# The "Segment Anything" foundation model achieves favorable brain tumor autosegmentation accuracy on MRI to support radiotherapy treatment planning.

Florian Putz[1,2], Johanna Grigo[1,2], Thomas Weissmann[1,2], Philipp Schubert[1,2], Daniel Höfler[1,2], Ahmed Gomaa[1,2], Hassen Ben Tkhayat[1,2], Amr Hagag[1,2], Sebastian Lettmaier[1,2], Benjamin Frey[1,2], Udo S. Gaipl[1,2], Luitpold V. Distel[1,2], Sabine Semrau[1,2], Christoph Bert[1,2], Rainer Fietkau[1,2], and Yixing Huang[1,2]

1. Department of Radiation Oncology, Universitätsklinikum Erlangen, Friedrich-Alexander-Universität Erlangen-Nürnberg, Erlangen, Germany
2. Comprehensive Cancer Center Erlangen-EMN (CCC ER-EMN), Erlangen, Deutschland

Author to whom correspondence should be addressed: Dr. Yixing Huang, Department of Radiation Oncology, Universitätsklinikum Erlangen, Friedrich-Alexander-Universität Erlangen-Nürnberg, yixing.huang@uk-erlangen.de

## Abstract

**Background:** Tumor segmentation in MRI is crucial in radiotherapy (RT) treatment planning for brain tumor patients. "Segment anything" (SA), a novel promptable foundation model for autosegmentation, has shown high accuracy for multiple segmentation tasks but was not evaluated on medical datasets yet.

**Methods:** SA was evaluated in a point-to-mask task for glioma brain tumor autosegmentation on 16744 transversal slices from 369 MRI datasets (BraTS 2020). Up to 9 point prompts were placed per slice. Tumor core (enhancing tumor + necrotic core) was segmented on contrast-enhanced T1w sequences. Out of the 3 masks predicted by SA, accuracy was evaluated for the mask with the highest calculated IoU (oracle mask) and with highest model predicted IoU (suggested mask). In addition to assessing SA on whole MRI slices, SA was also evaluated on images cropped to the tumor (max. 3D extent + 2 cm).

**Results**: Mean best IoU (mbIoU) using oracle mask on full MRI slices was 0.762 (IQR 0.713–0.917). Best 2D mask was achieved after a mean of 6.6 point prompts (IQR 5–9). Segmentation accuracy was significantly better for high- compared to low-grade glioma cases (mbIoU 0.789 vs. 0.668). Accuracy was worse using MRI slices cropped to the tumor (mbIoU 0.759) and was much worse using suggested mask (full slices 0.572). For all experiments, accuracy was low on peripheral slices with low tumor area (mbIoU, <300 mm²: 0.537 vs. ≥300: 0.841). Stacking best oracle segmentations from full axial MRI slices, mean 3D DSC for tumor core was 0.872, which was improved to 0.919 by combining axial, sagittal and coronal masks.

**Conclusions:** The Segment Anything foundation model, while trained on photos, can achieve high zero-shot accuracy for glioma brain tumor segmentation on MRI slices. The results suggest that Segment Anything can accelerate and facilitate RT treatment planning, when properly integrated in a clinical application.

# I. Introduction

Tumor segmentation in MRI constitutes the essential initial phase in radiotherapy treatment planning for patients with brain tumors [1,2]. Owing to the imaging heterogeneity of tumors, this indispensable image analysis step still predominantly relies on time-consuming manual contouring by physicians [2]. The substantial time demand of manual contouring restricts the accessibility of precise target volume definition, particularly in countries with limited resources [3]. Deep neural networks, such as U-net variants, have demonstrated exceptional accuracy in tumor autosegmentation within research studies [4,5]. However, implementing these models for specific tumor types and unique imaging data typically necessitates large, dedicated, expert-segmented training datasets that are representative for the data on which the model will be applied on later. Since out-of-distribution errors and poor generalizability can pose risks to treatment planning [6], automatic deep learning-based tumor autosegmentation has not yet experienced widespread clinical adoption. Furthermore, conventional deep learning autosegmentation models generally lack the capacity for expert-model interaction, which is necessary to efficiently perform corrections and adaptations to the model output under the guidance of clinical experts.

"Segment Anything" is a novel promptable foundation model for autosegmentation that has been trained on 11 million 2D photographs and over 1 billion segmentations masks [7]. The model enables promptable autosegmentation, in which the model output masks can be controlled using a variety of input prompts, such as foreground and background points, input boxes and masks, as well as text prompts. The "Segment Anything" model comprises a prompt encoder and an image encoder, both responsible for encoding the prompt and image input, and a mask decoder that connects to both encoders to predict valid segmentation masks. Interestingly, "Segment Anything" has been designed to predict three masks per prompt input to be able to deal with the common ambiguity of input prompts [7].

"Segment Anything" has shown high zero-shot autosegmentation performance for a large variety of 2D images from many different categories, including plant photography, traffic cameras, underwater scenes and microscopy; however, its performance on medical datasets still remains unexplored [7].

In the present manuscript, to the best of our knowledge, we investigate the "Segment Anything" foundation model for brain tumor autosegmentation on MRI datasets for the first time. Our work has a dual objective: first, we evaluate the segmentation accuracy of the "Segment Anything" model for treatment planning in glioma patients for the purpose of an expert-interactive workflow. Second, by assessing the zero-shot performance on brain tumor MRI datasets, we aim to further examine the generalizability of the "Segment Anything" foundation model for datasets that significantly deviate from its original training set.

Contributions:
In this manuscript, we make three core contribution. 1) We establish that "Segment Anything" can attain high segmentation accuracy for brain tumor MRI datasets in a point-to-mask setting. 2) We consequently demonstrate that "Segment Anything" effectively generalizes to brain tumor MRI datasets, achieving a segmentation accuracy comparable to that observed in 2D photographs on which it was previously evaluated. 3) We identify challenges encountered when applying "Segment Anything" to tumor segmentation in MRI datasets and identify strategies to address them, which are also applicable within the context of clinical implementation.

# II. Methods

Brain Tumor MRI Dataset:
In this study, we utilize the open BraTS 2020 glioma segmentation MRI dataset for evaluation purposes [8]. BraTS is a renowned glioma segmentation challenge in which research groups compete using their segmentation models. The BraTS 2020 segmentation challenge encompasses a training dataset consisting of 369 volumetric MRI studies and a validation set comprising 125 studies. Each study includes a volumetric pre-contrast and contrast-enhanced T1-weighted MRI sequence (T1ce), a T2-weighted sequence, and a T2-FLAIR weighted sequence [9]. As only the training dataset provides ground truth segmentation masks, rendering it suitable for an automatic evaluation of a point-to-mask task analogous to the methodology employed in the "Segment Anything" paper, we exclusively utilize the training dataset for our present evaluation [7].
The BraTS 2020 training set contains 259 datasets of patients with high-grade glioma and 110 datasets of low-grade gliomas. All datasets in the BraTS 2020 training set are already provided as spatially normalized, skull-stripped, and resampled to a common voxel grid of 240×240×155 with 1 mm-isotropic resolution. The ground truth segmentation masks comprise three classes: necrotic tumor core, contrast-enhancing tumor, and peritumoral edema. For evaluation purposes, these classes are combined in the BraTS challenge into tumor core (necrotic core and contrast-enhancing tumor), whole tumor (tumor core + peritumoral edema), and enhancing tumor [5,9]. To assess "Segment Anything" for its potential in supporting interactive clinical treatment planning and because the model is limited to a single image input, we evaluate segmentation accuracy using a single MRI sequence as input. As tumor core frequently represents the only relevant tumor compartment for clinical treatment planning [1] and as edema in the BraTS 2020 dataset was defined using multiple sequences, we exclusively examine tumor core segmentation on contrast-enhanced T1-weighted sequences.

Evaluation of autosegmentation performance:
We assess the "Segment Anything" segmentation performance on 2D MRI slices using a dedicated automated evaluation pipeline implemented in Python for 3D Slicer (v. 5.22) [10]. Inference was executed locally with the "Segment Anything" version 1.0 program code and ViT-H model weights on a workstation equipped with an NVIDIA RTX A6000 GPU, featuring 48 GB GPU memory (Python 3.9.10, Pytorch 1.10.1 with CUDA 11.1). MRI voxel intensities were normalized to a range between 0 and 255 by dividing them by the maximum intensity of each 3D dataset and subsequently multiplying by 255. Since "Segment Anything" anticipates a 3-channel image input, the rescaled intensities were replicated across all three channels. The automated point-to-mask autosegmentation performance evaluation involved the following steps:

1. Each transversal 2D MRI slice containing tumor core voxels was evaluated.
2. The initial point prompt was positioned at the tumor center on the respective slice (highest distance transform), and subsequent points were placed at the center of the set difference between the ground truth segmentation and the mask prediction, with a maximum of nine points being placed.
3. If the ground truth segmentation area exceeded the mask prediction, a foreground point was placed at the center of the set difference (ground truth – predicted mask); otherwise, a background point was placed at the center of the set difference (predicted mask – ground truth).
4. For each iteration, the Intersection over Union (IoU) between the ground truth mask and the predicted mask was calculated.

The best achieved IoU per slice was evaluated as the primary performance metric, reflecting an expert-interactive workflow in which the expert would retain the optimal mask prediction. To address input prompt ambiguity, "Segment Anything" produces three masks per prediction, along with an IoU estimate for each mask. We assessed two types of mask predictions: 1) the mask with the highest predicted IoU (suggested mask) and 2) the mask with the highest calculated IoU relative to the ground truth (oracle mask). In this evaluation, we primarily utilized the oracle mask, as it best represents an expert-interactive workflow where the expert selects the most accurate mask prediction.

In addition to evaluating "Segment Anything" on whole MRI slices, we also assessed model performance on cropped images using a cuboid region-of-interest (ROI) fitted to the 3D tumor extent with a 2 cm margin. For each experiment and both oracle and suggested masks, we evaluated a total of 16,744 2D transversal MRI slices. To facilitate comparability with 3D segmentation methods, 2D slice-based predictions were stacked, and the volumetric Dice score was calculated relative to the ground truth. Lastly, 3D segmentations from stacked slice-based predictions from transversal, coronal, and sagittal slices were combined using majority voting to provide an indication of the expected performance in an expert-interactive workflow where multiple slice orientations could be employed.

Statistical analyses were performed using SPSS (version 21.0.0.2, IBM) and R (version 3.5.2, R Project for Statistical Computing), with graphs created using SPSS and GraphPad Prism (version 9.5.0, GraphPad Software). The optimum threshold for the amount of tumor voxels per MRI slice in regard to segmentation accuracy was calculated using maximally selected rank statistics (R package maxstat, Wilcoxon statistics, p value adjusted for multiple testing). Differences between paired experiments were compared using a Wilcoxon signed-rank test, while unpaired experiments employed a Wilcoxon rank-sum test. P-values < 0.05 were deemed statistically significant.

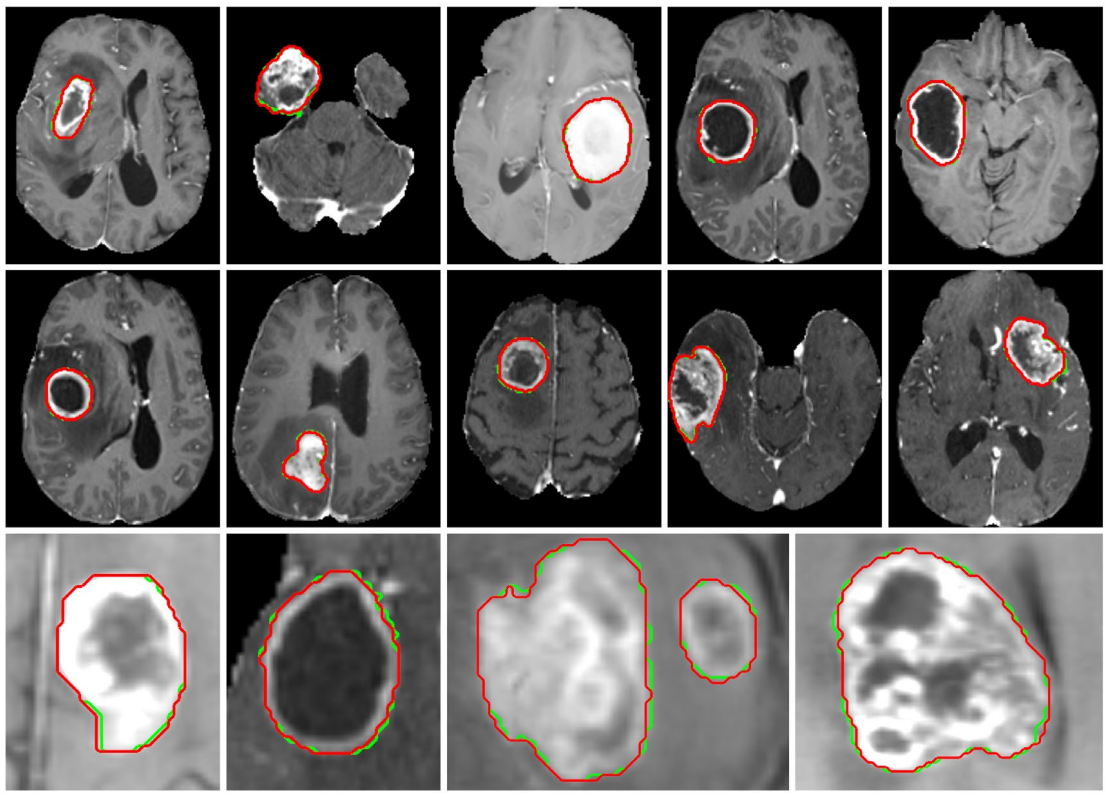

**Fig. 1.** Exemplary "Segment Anything" glioma autosegmentations on contrast-enhanced T1 sequences from the BraTS 2020 dataset. Green: ground truth segmentation; Red: Best "Segment Anything" oracle mask autosegmentation obtained with up to 9 automated centered point prompts. Bottom row: enlarged view.

# III. Results

Evaluation on whole MRI slices:

A total of 369 MRI datasets and 16,744 MRI slices were evaluated (Figure 1). Out of all 16,744 MRI slices evaluated; 13,035 slices were images of high-grade gliomas (HGGs) and 3,709 were images of low-grade gliomas (LGGs). Mean best Intersection over Union (IoU) per slice for oracle mask was 0.762 (interquartile range [IQR], 0.713 – 0.917). Interestingly, mean best IoU was significantly better for HGG compared to LGG cases: 0.789 (IQR, 0.770 – 0.923) vs. 0.668 (IQR, 0.541 – 0.855) ($p < 0.001$, Figure 2). Mean number of point prompts required to obtain the best mask prediction was 6.6 (IQR, 5 – 9).

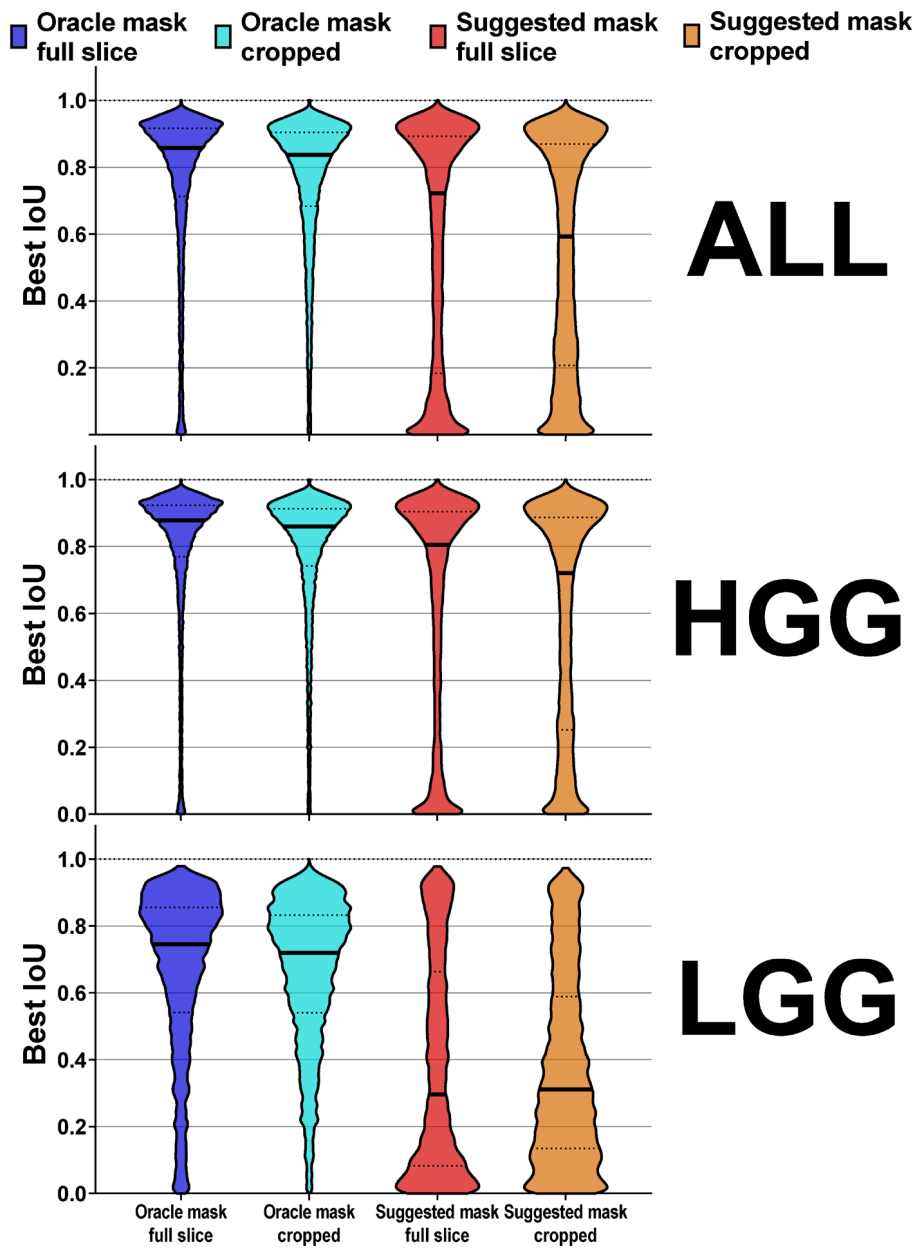

**Fig. 2.** Mean best Intersection over Union (IoU) obtained with "Segment Anything" point-to-mask autosegmentation in the BraTS 2020 training dataset (n = 369 contrast-enhanced T1 datasets, 16744 MRI slices). Blue: oracle mask on full MRI slices, cyan: oracle mask on MRI slices cropped to the 3D tumor extent + 2 cm, red: model suggested mask on full MRI slices, orange: model suggested mask on cropped slices. Top row: results for all datasets, middle row: results for high-grade glioma (HGG) datasets, bottom row: results for low-grade glioma (LGG) datasets. Note: Decrease in accuracy with LGG compared to HGG cases.

Change in Accuracy with the Number of Input Point Prompts:
Single point-to-mask segmentation IoU was 0.586 (IQR, 0.314 – 0.864) for oracle mask. For HGGs, single point accuracy was 0.640 (IQR, 0.465 – 0.880) but only 0.397 (IQR, 0.110 – 0.669) for LGGs (p < 0.001). Mean IoU using oracle mask consistently improved for HGG as well as LGG cases with the number of point prompts (Figure 3).

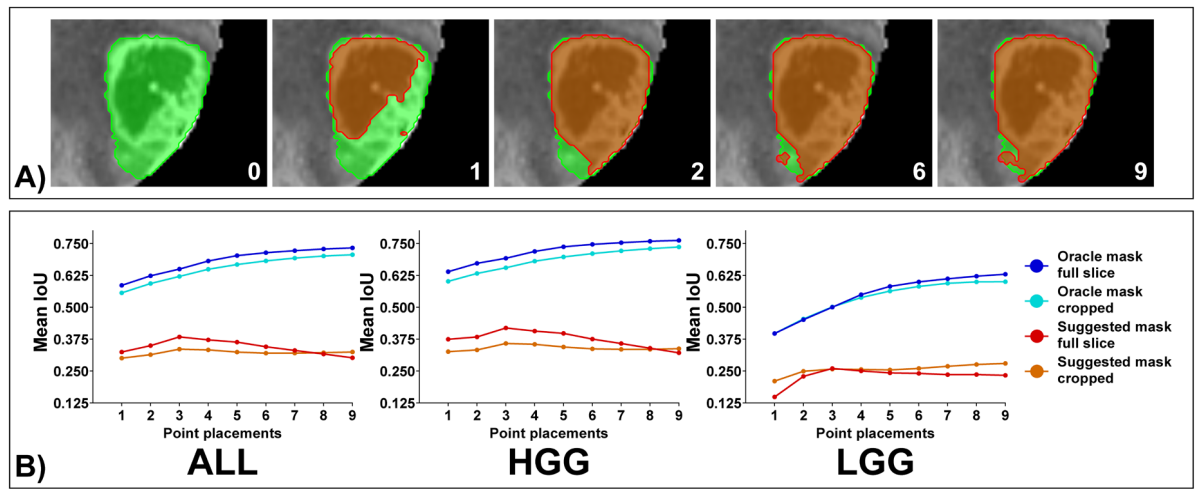

**Fig. 3.** "Segment Anything" segmentation accuracy in relation to the number of prompted points. A) An example case illustrating the increased similarity between the ground truth segmentation and the "Segment Anything" mask prediction as the number of point placements increases (green: ground truth segmentation, red: autosegmentation [oracle mask, full MRI slices]). B) Mean IoU for 1 to 9 point prompts for the entire dataset (ALL, left), high-grade (HGG, middle), and low-grade glioma cases (LGG, right). Note: The IoU improvement is observed with the increase in the number of point prompts for the oracle mask but not for the model-suggested mask.

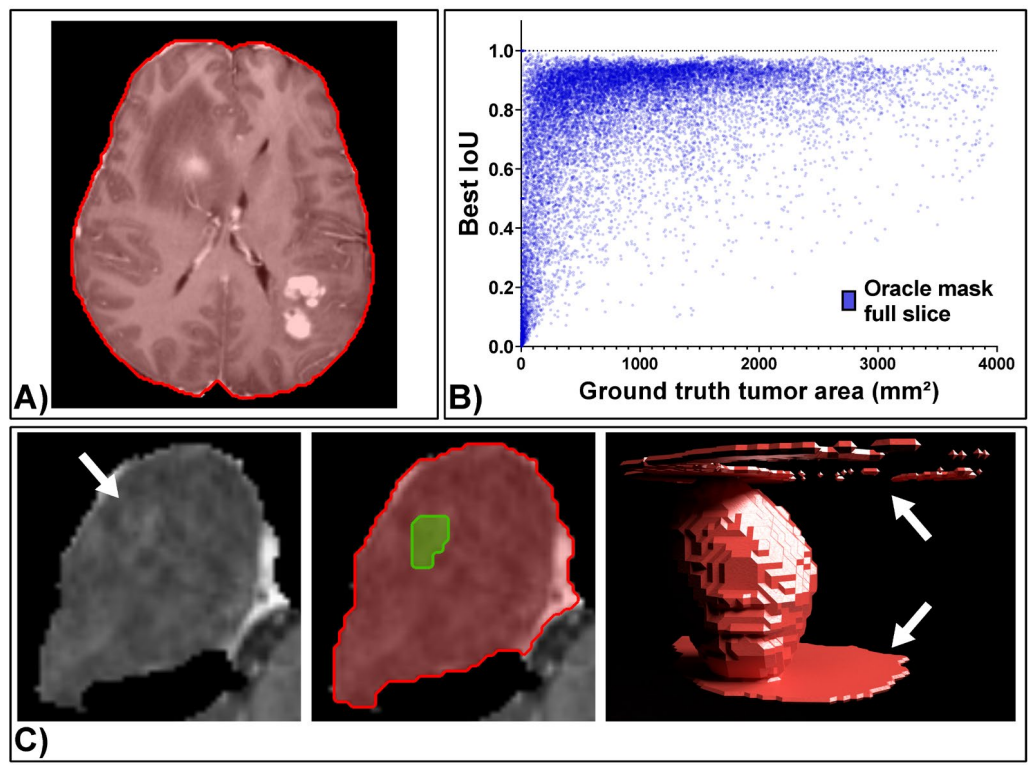

Fig. 4. Common autosegmentation errors observed with "Segment Anything" on T1 contrast-enhanced glioma MR images. A) Whole patient/brain circumference is segmented by "Segment Anything" instead of the tumor, because of prompt ambiguity. B) Scatter plot of best IoU as a function of ground truth tumor area. Note: Low accuracy segmentations are primarily observed for MRI images with few tumor voxels. C) Example case illustrating problems with segmentation on peripheral slices with few tumor voxels after single point placement.

Common Segmentation Errors:
Figure 4 shows examples of common segmentation errors. Because of prompt ambiguity, "Segment Anything" sometimes segmented the whole patient / brain circumference instead of the tumor. This was especially observed, for single point prompts, for model suggested instead of oracle mask, and at peripheral slices with few tumor voxels. Segmentation errors were mostly observed on peripheral slices, that only contained few tumor voxels (Figure 4 B and C). Figure 4 B illustrates that low segmentation accuracy clustered to slices with few tumor voxels and that there was a correlation between best IoU and the ground truth tumor area (Spearman $\rho = 0.495$, $p < 0.001$). A minimum tumor area of 300 mm² per slice was the optimal threshold separating slices with high and low segmentation accuracy by maximally selected rank statistics. Mean best IoU for oracle mask was 0.841 (IQR, 0.804 – 0.926) for a minimum tumor area of 300 mm², whereas it was only 0.537 (IQR, 0.244 – 0.821) for a tumor area < 300 mm² (adjusted $p < 0.001$). Single point to mask mean IoU was 0.670 (IQR, 0.499 – 0.883) vs. 0.346 (IQR, 0.018 – 0.699) for tumor areas of ≥ and < 300 mm², respectively. The highest single point to mask mean IoU for oracle mask was observed for MRI image slices of high-grade gliomas with a tumor area of ≥ 300 mm² (0.735).

Evaluation on MRI slices cropped to the tumor region:
Since segmentation errors involving the entire patient circumference were frequently observed, we also assessed "Segment Anything" on image patches cropped to the 3D tumor circumference with an additional 2 cm margin. This could easily be integrated into an expert interactive workflow and would correspond to the expert zooming-in on the tumor region. However, contrary to expectations, cropping did not enhance segmentation accuracy but rather diminished it (Figure 2, 3). The mean best IoU was 0.762 (IQR, 0.713 – 0.917) without cropping and 0.759 (IQR, 0.683 – 0.905) with cropping ($p < 0.001$). Furthermore, the oracle mask consistently displayed poorer performance when applying cropping as opposed to not cropping across all numbers of input point prompts (Figure 3B). A similar decline in performance was observed for whole-slice image input with an additional, equally sized box prompt, and even for a box prompt precisely fitted to the 3D tumor extent. Box prompts tailored to individual 2D slices were not examined, as combining box and point prompts on each individual slice was considered too time-consuming for a clinical workflow.

Results for Model Suggested Mask Prediction:
The model-suggested mask consistently exhibited substantially lower performance compared to the oracle mask (Figure 2, 3). The mean best IoU for the oracle mask without cropping was 0.762 (IQR, 0.713 – 0.917), while for the model-suggested mask, it was only 0.572 (0.184 – 0.892, $p < 0.001$). Intriguingly, no consistent improvement for the model-suggested mask was observed with increasing number of input point prompts (Figure 3).

3D Segmentation Accuracy with and without Combination of Multiple Slice Orientations
To evaluate the accuracy of "Segment Anything" point-prompt-based segmentation in comparison to other methods published for glioma autosegmentation, we calculated the volumetric Dice similarity score by stacking the "Segment Anything" 2D masks. For all 369 BraTS training datasets, the mean volumetric tumor core Dice score achieved by stacking the best oracle mask 2D segmentations from transversal slices was 0.872 (IQR, 0.843 – 0.941). As segmentation errors were predominantly observed for peripheral slices (see above), the mean volumetric Dice score could be further improved to 0.919 (IQR, 0.903 – 0.956, p for difference < 0.001) by combining "Segment Anything" 2D masks from transversal, coronal, and sagittal slice orientations (Figure 5). This approach would correspond to a clinical expert utilizing "Segment Anything" across multiple slice orientations in an interactive workflow.

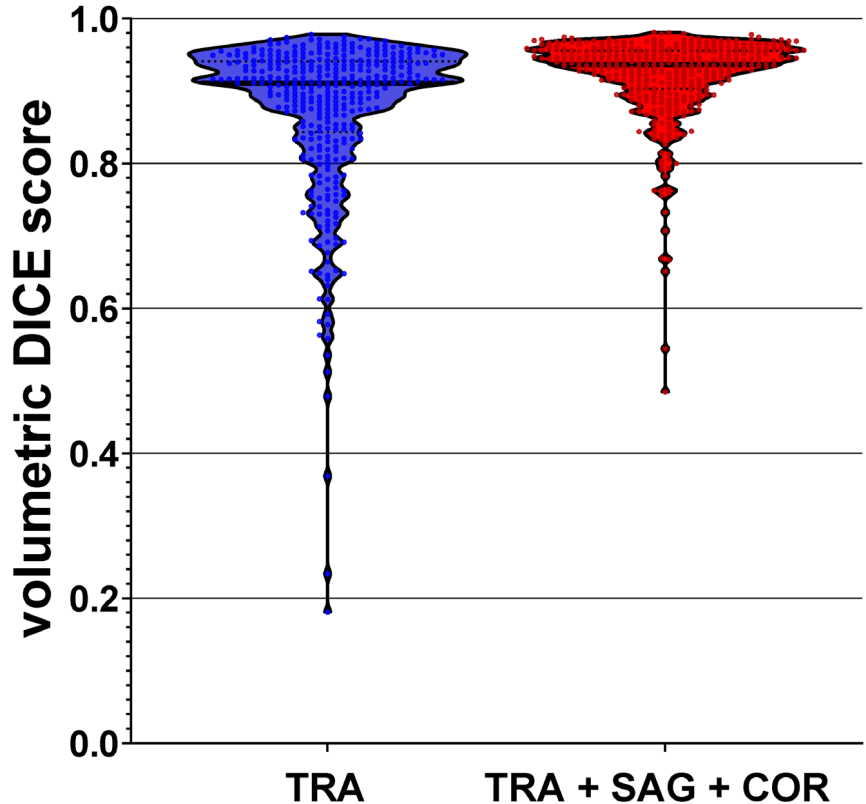

**Fig. 5.** Volumetric Dice score achieved by stacking "Segment Anything" 2D segmentation masks (best oracle mask prediction per slice, full MRI slices). Left: "Segment Anything" segmentation using only transversal slices. Right: "Segment Anything" segmentation employing transversal, sagittal, and coronal slices, combined via simple majority voting, which corresponds to a potential interactive clinical workflow where an expert can utilize all three slice orientations. Note: significant improvement in 3D segmentation accuracy by combining all three slice orientations.

## IV. Discussion

"Segment Anything" achieved high zero-shot accuracy in this point-to-mask segmentation evaluation in reference to state-of-the art fully automatic glioma autosegmentation methods. The volumetric Dice score for tumor core segmentation by stacking best oracle mask predictions from transversal slices was 0.872 and even reached 0.919 by combining 2D masks from transversal, sagittal and coronal slice orientations. In contrast, the winners of the BraTS 2020 segmentation challenge, Isensee et al. using an optimized nnU-net implementation reported a tumor core Dice score of 0.8506 in the BraTS 2020 validation set. [5,8] The two evaluations must not be directly compared, as the tasks were completely different (fully automatic 3D segmentation after supervised learning vs. zero-shot point-to-2D mask) and the evaluation dataset differed (BraTS 2020 training dataset vs. validation dataset). However, the juxtaposition illustrates that "Segment Anything" could indeed achieve a high segmentation accuracy in an expert-interactive workflow.

The generalizability and zero-shot performance of "Segment Anything" to brain tumor MRI datasets can be interpreted as remarkable, as it achieved a practically useful level for clinical tasks. The difference between tumor segmentation in MRI datasets and object segmentation in 2D photographs should be large. Image slices from 3D MRI datasets are acquired based on completely different physical principles than conventional 2D photographs [11]. Moreover, glioma brain tumors are no discrete objects in three-dimensional space but are highly infiltrative tumors that are to some extent continuous with their surroundings [12-14]. The segmentation accuracy observed for "Segment Anything" in brain tumor MR images was slightly worse than for 2D photographs reported by the "Segment Anything" authors [7]. "Segment Anything" was evaluated by the authors across 23 datasets with 2D photographs ranging from plant images, aerial photographs, traffic camera footage, cooking photos to images of recycling waste and underwater scenes [7]. The oracle mask "Segment Anything" single point to mask mean segmentation accuracy was slightly below an IoU of 0.750 (value obtained from plot), whereas it was 0.586 in the present evaluation. However, the mean single point to mask accuracy, was larger for high-grade than low-grade gliomas, was dragged down by peripheral slices with low tumor area and consequently reached 0.735 on high-grade glioma MRI slices with at least 300 tumor voxels. Moreover, the segmentation accuracy increased consistently for oracle mask predictions with an increasing number of input point prompts.

To be able to handle prompt ambiguity, "Segment Anything" makes 3 mask predictions per input image [7]. In this manuscript, we evaluated the objectively best mask prediction (oracle mask) that had the best calculated IoU as the main evaluation metric. In addition, we also evaluated the most confident mask prediction (suggested mask) that had the highest model predicted IoU. In stark contrast to the oracle mask accuracy, segmentation performance for the model suggested mask was very low and did not consistently improve with the number of input point prompts (Figure 3). This finding could indicate, that the "Segment Anything" IoU predictions do not perform well in the context of medical datasets and segmentation tasks.

The reliance on oracle mask predictions could be seen as a disadvantage for implementation into a clinical application, as for conventional 2D photographs this would correspond to the interacting user having to evaluate all 3 mask predictions for every image. However, because of the redundancy of high-resolution medical 3D datasets that change very little from one slice to another, the finalized segmentation mask of the previous slice is a valid approximation for the ground truth mask of each slice. Thus, with sequential slice-wise segmentation the mask prediction most similar to the previous slice segmentation can be automatically selected to achieve a similar performance as obtained with the oracle mask prediction in this evaluation. Therefore, the evaluation performed in the current work can readily be integrated in a clinical workflow. To provide a practical example for a clinical implementation of "Segment Anything" and to facilitate further clinical assessments, we have developed and share a corresponding segmentation extension for the open-source software 3D Slicer* [10].

Because of the heterogeneity of tumors and imaging data, deep learning tumor autosegmentation has not seen wide-spread clinical adoption for treatment planning yet. Dedicated deep learning autosegmentation models, trained for specific tumor types in specific medical imaging datasets, however, have achieved high accuracy in the context of scientific studies [4,5,15,16]. For brain tumors, these models have shown very promising performance for glioma segmentation, but also for brain metastases and vestibular schwannomas [5,15,16]. For brain metastases, Lu et al. were able to demonstrate in a multireader study that assistance by a deep learning ensemble autosegmentation model (3D U-Net + DeepMedic), could improve brain metastases detection sensitivity from 82.6% to 91.3% and the inter-reader Dice score form 0.86 to 0.90 compared to manual expert evaluation alone [17]. 3D and 2D U-net variants have shown particular high autosegmentation performance and have seen wide-spread evaluation and adaption for a vast number of biomedical segmentation tasks [4,18]. As many proposed U-net improvements showed inconsistent benefit outside their original datasets as well as limited comparability due to overoptimization, nnU-net was proposed [4]. NnU-net or "No-new-U-net" has become a standard for autosegmentation and has shown state-of-the-art performance across multiple biomedical segmentation tasks [4]. Rather than being optimized for one particular dataset, NnU-net exhibits broadly applicable design choices and automatically adapts to a dataset at hand, which includes configuring multiple network hyperparameters. Moreover, NnU-net also features an extensive on-the-fly image augmentation pipeline and 3D as well as 2D U-net architectures [4]. Because of the great success of dedicated autosegmentation models trained in a supervised fashion for specific medical segmentations tasks. More wide-spread introduction of these models in clinical radiotherapy treatment planning is only a question of time.

\* program code will be published after full manuscript acceptance.

However, large promptable foundation models for autosegmentation like "Segment Anything" that show high zero-shot accuracy and can generalize to a vast amount of segmentation problems should still have an important role to play. Dedicated deep learning autosegmentation models like U-net variants trained in conventional supervised fashion, usually can only be applied to a very narrow range of tumor types and image datasets, that correspond to a preexisting training dataset [4]. While such dedicated models can be created for frequently encountered tumor and image dataset types like e.g., brain metastases in contrast-enhanced 3D-T1w gradient echo sequences, a substantial proportion of segmentation problems will probably remain, for which the generation, evaluation and regulatory approval of dedicated segmentation models in the present form won't be economically feasible. Moreover, dedicated deep learning autosegmentation models showing high segmentation accuracy at a cohort level, can still fail for very specific patient cases and unseen image datasets, because of still limited generalizability, making expert review and correction of autosegmentations mandatory [6]. Finally, slice-based 2D promptable autosegmentation models like "Segment Anything" could also greatly accelerate the creation of annotated training datasets for supervised training of dedicated deep learning autosegmentation models.

Most deep learning tumor autosegmentations necessitate some degree of expert correction to obtain the final segmentation for treatment planning. Since dedicated deep learning autosegmentation models do not allow for user interaction, correcting autosegmentations typically involves fully manual, slice-wise adjustments to the segmented tumor boundary, which is highly inefficient [19]. Drawing an analogy, this sequence of tumor autosegmentation and manual correction is akin to traveling from one city to another by plane, only to walk the remaining distance from the airport to the final destination by foot. As "Segment Anything" introduces a novel promptable image segmentation task, it is ideally suited for expert-interactive segmentation tasks. "Segment Anything" allows for multiple types of input prompts, which would allow clinical experts to make rapid adaptions and corrections to segmentations. The present evaluation clearly demonstrates this adaptation of mask predictions towards the ground truth segmentation (e.g., see Figure 3). In addition to foreground and background point prompts, which were mainly evaluated in the present work, "Segment Anything" also allows for box, mask and text input prompts [7]. Additional box prompts were not helpful in the present evaluation and therefore were not further pursued. However, box prompts should be enabled in clinical implementations for an expert interactive workflow to give the user further freedom for interaction with the autosegmentation model, which should be helpful for certain segmentation problems. Finally, mask and text prompts were not evaluated in the present manuscript. While introducing text prompts for medical segmentation problems is highly interesting, we did not include it in this initial evaluation, as text prompts would be very difficult to implement in an automated evaluation and as we expected the CLIP-based text prompt implementation to perform poorly with medical nomenclature on MRI datasets. Finally, mask-based prompts were not evaluated in the present work, as currently mask input in "Segment Anything" is limited to low-resolution logit masks and drawing input masks for individual slices would not accelerate clinical treatment planning in a meaningful way [7]. However, mask input in the context of medical 3D dataset segmentation is interesting and could, e.g., allow to propagate 2D segmentations across multiple slices.

The fact that "Segment Anything" only allows for 2D segmentations can be seen as a disadvantage for segmenting 3D medical datasets. While certainly, a similar promptable foundation model for 3D segmentation in volumetric medical datasets would be greatly beneficial for treatment planning, 2D autosegmentation has the advantage that every image slice is segmented under expert guidance in an interactive way and fits very well into current applications for radiotherapy treatment planning. Aside from facilitating and accelerating treatment planning, the high 2D segmentation accuracy of "Segment Anything" could also support diagnostic applications, like automatic perpendicular diameter measurements that currently form the foundation of response assessment in gliomas [20].

Although the zero-shot segmentation performance and generalizability of "Segment Anything" for clinical applications, such as glioma brain tumor autosegmentation in MRI datasets, are undoubtedly remarkable, additional performance improvements may be achieved by fine-tuning "Segment Anything" on medical datasets. The present evaluation supports this assumption, as common segmentation errors included whole brain area segmentation and inaccurate IoU predictions for the three output masks.

Groundbreaking breakthroughs in deep learning models like GPT-4 and "Segment Anything", which demonstrate unparalleled generalization abilities, have resulted from massive training datasets [7,21]. "Segment Anything" was trained on 11 million images and over one billion masks [7]. Although details about GPT-4's training data remain undisclosed, it is likely to encompass a significant portion of the internet's available text data [21].

Regrettably, extensive medical datasets of a similar scale, which could substantially benefit humanity, are not currently available. The primary obstacles to similar progress in the medical domain include overly restrictive data protection regulations that impede scientific advancement without significantly enhancing individual privacy [22]. Policymakers should, therefore, establish appropriate framework conditions to facilitate, simplify, and incentivize sharing and pooling of medical datasets. Furthermore, philanthropic organizations could contribute to the development of large medical datasets by licensing medical data, akin to the photographic data licensed by the "Segment Anything" initiative. In the meantime, it is advantageous that large foundation models like "Segment Anything," trained on vast non-medical datasets, can be meaningfully transferred to the medical domain, potentially including further fine-tuning on smaller medical datasets.

Conclusions:

The "Segment Anything" promptable segmentation foundation model demonstrated high zero-shot accuracy for glioma brain tumor point-to-mask autosegmentation on MRI slices. These findings indicate that "Segment Anything" can accelerate and facilitate RT treatment planning, when properly integrated in a clinical application. Furthermore, the results demonstrate the ability of "Segment Anything", as a large promptable foundation model, to generalize effectively to unseen data from a novel domain.